# Etude cognitive des processus de construction d'une requête dans un système de gestion de connaissances médicales

N. Chaignaud\*, V. Delavigne\*\*, M. Holzem\*\*, J-Ph. Kotowicz\*
et A. Loisel\*

*\*Laboratoire LITIS - EA 4108*
*Place Emile Blondel - BP 08 - 76131 Mont-Saint-Aignan Cedex*

*\*\*Laboratoire LiDiFra*
*7, rue Thomas Becket - 76821 Mont-Saint-Aignan Cedex*

RÉSUMÉ. *Cet article décrit le projet Cogni-CISMeF qui propose un module de dialogue Homme-Machine à intégrer dans le système d'indexation de connaissances médicales CISMeF (Catalogue et Index des Sites Médicaux Francophones). Nous avons adopté une démarche de modélisation cognitive en procédant à un recueil de corpus de dialogues entre un utilisateur (jouant le rôle d'un patient) désirant une information médicale et un expert CISMeF affinant cette demande pour construire la requête. Nous avons analysé la structure des dialogues ainsi obtenus et avons étudié un certain nombre d'indices discursifs : vocabulaire employé, marques de reformulation, commentaires méta et épilinguistiques, expression implicite ou explicite des intentions de l'utilisateur, enchaînement conversationnel, etc. De cette analyse, nous avons construit un modèle d'agent artificiel doté de capacités cognitives capables d'aider l'utilisateur dans sa tâche de recherche d'information. Ce modèle a été implémenté et intégré dans le système CISMeF.*

ABSTRACT. *This article presents the Cogni-CISMeF project, which aims at improving medical information search in the CISMeF system (Catalog and Index of French-language health resources) by including a conversational agent to interact with the user in natural language. To study the cognitive processes involved during the information search, a bottom-up methodology was adopted. Experimentation has been set up to obtain human dialogs between a user (playing the role of patient) dealing with medical information search and a CISMeF expert refining the request. The analysis of these dialogs underlined the use of discursive evidence: vocabulary, reformulation, implicit or explicit expression of user intentions, conversational sequences, etc. A model of artificial agent is proposed. It leads the user in its information search by proposing to him examples, assistance and choices. This model was implemented and integrated in the CISMeF system.*

MOTS-CLÉS : *Interaction, dialogue Homme-Machine, modélisation cognitive, analyse de corpus.*

KEYWORDS: *Interaction, human-computer dialog, cognitive modeling, corpus analysis.*

**1. Introduction**

À notre connaissance, aucun moteur de recherche sur l'internet ne cherche à prendre en compte réellement les intentions de l'utilisateur, ni ne lui permet de décrire ce qu'il recherche sous la forme d'un véritable dialogue Homme-Machine (H-M). Pourtant cette forme d'interface présente de nombreux avantages dans la mesure où elle tente de s'appuyer sur la réalité des pratiques communicatives, les stratégies discursives et les variations langagières, en intégrant les requêtes antérieures qui contextualisent la requête courante. Cependant, pour enrichir l'interaction H-M, il faut chercher à mettre en évidence non seulement les facteurs externes qui déterminent les comportements face à la machine, mais aussi les phénomènes susceptibles d'expliquer l'élaboration de la requête par l'utilisateur.

Notre but est de concevoir un système de dialogue H-M pour la recherche d'informations capable d'interagir avec l'utilisateur en langue naturelle et de lui proposer des stratégies coopératives. La machine tente pour cela d'analyser l'objectif de l'utilisateur et lui propose des solutions pour évaluer l'état de la recherche d'informations. Elle présente des exemples, des aides, des corrections ou des clarifications sous forme de choix. Elle accompagne l'interlocuteur jusqu'à une solution en élargissant son but initial si nécessaire. Un tel dialogue propose à la fois des éléments pour expliciter le cheminement de l'utilisateur et examine avec lui les mots-clés à employer. Le système teste plusieurs requêtes en utilisant des critères permettant d'orienter la recherche dans la direction souhaitée par l'utilisateur. De plus, le dialogue en langue naturelle permet de constituer un terrain partagé et offre des moyens d'établir ce terrain commun, chose impossible à mettre en place dans une navigation hypermédia classique.

Le domaine d'application de notre étude est le projet Cogni-CISMeF (financé par le Programme Interdisciplinaire TCAN du CNRS) (Kotowicz *et al.*, 2007) qui vise à intégrer un module de dialogue H-M dans le système d'indexation de connaissances médicales CISMeF (Catalogue et Index des Sites Médicaux Francophones).

Contrairement aux démarches a prioristes qui cherchent à typifier et réduire les intentions de l'humain, notre approche interdisciplinaire a pour but de concevoir un système apte à s'adapter à un interlocuteur humain. De ce point de vue, les communications H-M ne peuvent être conçues que grâce à l'étude des interactions Homme-Homme (H-H). Nous sommes certes conscients que le dialogue H-M est différent du dialogue H-H, mais l'analyse d'interactions H-H devrait toutefois permettre d'obtenir des indices indispensables à la mise en place d'un dialogue H-M, notamment en ce qui concerne la structure du dialogue.

Pour atteindre un tel objectif, nous avons procédé à un recueil de corpus de dialogues entre un utilisateur (jouant le rôle d'un patient) désirant une information médicale et un expert CISMeF affinant cette demande pour construire la requête. L'analyse de la structure des interactions ainsi obtenues et l'étude d'un certain nombre d'indices discursifs nous ont permis de dégager les principales caractéristiques de ces dialogues : vocabulaire employé, marques de reformulation, expression implicite ou explicite des intentions de l'utilisateur, enchaînements conversationnels, etc. Notre modèle de dialogue H-M a été conçu à partir de l'étude de ce corpus et nous avons montré que le système de dialogue GoDIS (Larsson, 2002) répond en partie à nos besoins. Ainsi, nous essayons de contribuer, modestement, à réduire le fossé entre les systèmes de dialogue H-M et la communication entre humains.

La méthodologie utilisée convoque une enquête sociolinguistique et mène à la constitution d'un corpus en prise sur les usages réels (enregistrements et retranscriptions de dialogues H-H concomitants à la recherche de documents médicaux sur CISMeF). L'analyse du corpus vise à comprendre la dynamique des interactions et à mettre en évidence leurs principales caractéristiques sans partir d'un modèle préalable du dialogue. Une première modélisation est issue de cette analyse, puis affinée en retournant au corpus par un mouvement de va-et-vient régulier entre analyse de plus en plus fine et modélisation qui se précise.

L'ensemble étant supérieur à la somme des parties, le fruit de la collaboration linguistes-informaticiens a permis de trouver un compromis entre opérationnalisation et usages réels. Du point de vue linguistique, cette méthodologie permet de questionner les phénomènes d'interprétation à l'œuvre dans l'interrogation d'une base de documents tout en montrant les limites du computationnalisme et, du côté informatique, de proposer un modèle étendu qui combine les atouts de plusieurs approches de dialogue H-M.

Cet article suit un plan linéaire à notre méthodologie : la section 2 expose un bref état de l'art sur les systèmes de dialogue ; la section 3 décrit le cadre expérimental permettant le recueil de notre corpus dont l'analyse est développée en section 4 ; la section 5 présente la modélisation de l'agent dialogique découlant de notre analyse avec son implémentation et un exemple de dialogue avec notre système ; enfin, la section 6 conclut cet article et donne quelques perspectives.

**2. Les systèmes de dialogue**

Nous nous concentrons ici sur les modèles conventionnels du dialogue qui utilisent la notion de tableau conversationnel. Nous détaillons plus spécifiquement une théorie fondée sur la sémantique des questions : la théorie *QUD (Question Under Discussion)* (Ginzburg, 1996). Cette théorie a débouché sur une modélisation informatique GoDIS (Larsson, 2002) englobant aussi certains principes issus de la planification (approche mixte).

**2.1. La théorie des questions en discussion (QUD)**

Le modèle QUD (Ginzburg, 1996) est une théorie conventionnelle du dialogue fondée sur une sémantique formelle servant de modèle explicatif pour la résolution des ellipses et de certaines présuppositions. L'originalité de ce modèle par rapport à ceux traitant des *questions* comme (Groenendijik et Stokhof, 1984) est de proposer une version structurée d'un tableau de conversations dirigé par la sémantique des questions en contexte. Le but de QUD est de déterminer très précisément les propriétés des couples *Qestions-Réponses (Q-R) (Issues* en anglais*)* et de montrer comment les questions et assertions en discussion enrichissent le tableau de conversation : que peut-on asserter ou demander à un instant donné ? En reprenant l'approche par jeux de dialogue ((Levin 1980), (Maudet 2001)), Ginzburg conçoit le dialogue en termes de *coups* dans un jeu. Cette approche nous semble doublement prometteuse. Tout d'abord, le jeu appelle une conception dynamique du dialogue, non prédéfinie préalablement comme pourrait l'être un scénario pouvant être rejoué à l'identique. L'issue d'un jeu ne saurait être prédéterminée. D'autre part, le jeu nécessite la participation pleine et entière du sujet engagé dans l'interaction. Il s'accompagne même, diraient les herméneutes (Gadamer 1976), d'un oubli de soi (qui explique entre autre le succès des jeux vidéos dont rêvent bon nombre de concepteurs de plate-forme de dialogues).

**2.2. Les dialogues fondés sur les *issues* et GoDIS**

Fondé sur QUD, GoDIS (Larsson, 2002) est un modèle de dialogue computationnel ne gardant qu'une sémantique simplifiée des questions. Il est possible de représenter trois types de question : interrogation totale, interrogation partielle ou interrogation parmi une liste de choix. Les Q-R sont intégrées dans une structure de plans de dialogue (plans composés d'une séquence d'actions abstraites appelées *actions de plan*). Les plans dans GoDIS représentent à la fois la tâche et le dialogue. GoDIS utilise la notion l'*état d'information (IS)*, concept proche du tableau de conversation, qui se décompose en deux parties :

```
Private
   Agenda : file d'actions
   Plan : pile de plans
   Believe : ensemble de propositions
   Not integrated moves : file d'actes de dialogue
Shared
   Com : ensemble de propositions
   Qud : pile de questions
   Issues : pile de questions
   Actions : pile d'actions
   Previous moves : file d'actes de dialogue
   Last utterance : (énonciateur : participant)
                    (coups : ensemble d'actes de dialogue)
```

La partie partagée permet de définir le tableau conversationnel mémorisant des informations partagées par les deux interlocuteurs. Des règles de mise à jour et de sélection permettent de contrôler l'IS.

Le principe d'accommodation donne de la souplesse au dialogue. Il permet de mettre ou remettre des questions en discussion de manière non triviale.

### 2.3. Discussion

Le concept de Q-R peut être considéré comme une spécialisation des jeux de dialogues (Hulstijn, 2000). Comme le souligne (Beveridge et Milward, 2000), on peut définir des relations intentionnelles entre jeux de dialogues ainsi que des relations sémantiques. Or GoDIS ne définit que la seule relation de dominance entre Q-R et cette relation n'est pas récursive.

D'autre part, aucune relation sémantique n'est définie et les différentes Q-R se succèdent par une relation de séquence implicite. Puisque aucune contrainte de *satisfaction-précédence* n'existe, GoDIS permet d'accommoder tout plan de dialogue à tout moment. Ce modèle bien que riche ne permet de traiter que des dialogues simples où chaque plan est défini comme une séquence simple d'actions ou de questions.

De plus, sauf pour les phénomènes interactionnels de clarification ou d'établissement du *terrain commun*, le dialogue ne peut s'éloigner de la tâche pour introduire une digression, une explication ou pour proposer des suggestions à l'utilisateur. (Caelen, 2003) propose la notion de stratégie de dialogue : le changement de stratégie vise à choisir la meilleure direction d'ajustement des buts à un moment donné. Deux types de stratégies nous intéressent :

– la *stratégie coopérative* vise à modifier le but courant pour s'adapter au but de l'interlocuteur. Dans GoDIS, cela revient à modifier dynamiquement les actions de plan dans la pile d'actions de plan (par exemple, proposer des suggestions en fonction de l'IS) tout en gardant le même but ;

– la *stratégie constructive* vise à abandonner provisoirement le but courant pour un nouveau but. Dans GoDIS, cela revient à abandonner la pile d'actions de plan pour un nouveau but. On peut ainsi introduire des digressions et ne pas suivre directement les séquences liées à la tâche.

Le principal avantage de GoDIS est de disposer d'un seul niveau de plan tout en étant très générique. Les actions de dialogue utilisées rendent la modélisation du dialogue proche de celle de la tâche. De plus, ce formalisme de plans permet de décrire des relations de dépendance entre questions et entre actions mais en assouplissant la notion de satisfaction-précédence. L'ordre des actions et questions n'est pas contraint au sein d'un même plan de dialogue.

### 3. Recueil de corpus

Comme le montre le courant de la linguistique de corpus et les travaux autour de l'IA en linguistique informatique ((Aussenac *et al.* 2003), (Rastier 2008)), un corpus est une construction élaborée dans un cadre précis. Il serait peu raisonnable de chercher à obtenir à travers un corpus l'exhaustivité des phénomènes qu'on cherche à étudier. L'étude d'un corpus relève d'une approche inductive et nécessite de faire appel à des connaissances extérieures pour compléter les données repérées. Le corpus se doit d'être caractéristique de l'ensemble à étudier (toute représentativité étant illusoire) en tentant de prendre en compte les différents types d'acteurs concernés, ainsi que les genres habituellement produits et interprétés au sein des sphères d'activité concernées.

Dans notre cadre, le corpus est considéré en tant que source de données qui doit nous permettre d'élaborer notre système de dialogue.

Nous avons mis en place une expérimentation afin d'obtenir des dialogues finalisés *in situ*. Deux membres du projet, après avoir été formés à CISMeF, jouent le rôle d'experts. Les questions proviennent, sur la base du volontariat, de membres du laboratoire LITIS (secrétaires, administrateur réseau, doctorants, enseignants-chercheurs). Pour éviter tout biais provenant d'une connivence entre membres d'un même laboratoire, seuls Valérie Delavigne et Alain Loisel (moins familiers avec les membres du laboratoire) ont réalisé cette expérimentation en jouant le rôle d'experts, l'intervention de deux expérimentateurs permettant de contraster les démarches.

Lors de chaque expérimentation, chaque expert se retrouve en tête à tête avec un interlocuteur et dispose d'un accès à CISMeF. L'expert mène la recherche et doit en même temps verbaliser tout ce qu'il est en train de faire. Les logs des requêtes sont récupérés par le système et les dialogues sont enregistrés afin d'être retranscrit. L'entretien se clôt lorsque la réponse satisfait le demandeur, ou qu'il semble bien qu'aucune réponse ne puisse être trouvée. Notre corpus est constitué des retranscriptions des 21 dialogues enregistrés lors de cette expérimentation.

D'autre part, parallèlement à cette expérimentation, nous avons demandé à Benoît Thirion, expert CISMeF, de répondre aux demandes formulées par les membres du LITIS pour obtenir les requêtes « optimales » correspondant à ces demandes. La verbalisation de la construction de ces requêtes a permis de souligner les différentes stratégies employées. Ces verbalisations enregistrées (mais non retranscrites) ont fourni une base de travail pour proposer une modélisation de la tâche de recherche d'informations.

## 4. Analyse du corpus

Le corpus retranscrit a été analysé sur cinq niveaux.

– Une analyse sociolinguistique a permis de mettre en évidence certains phénomènes langagiers : précautions oratoires, mise en situation, tentatives d'influencer l'interlocuteur, etc.

– Une analyse des sous-dialogues fait apparaître la structure globale du dialogue comme succession d'étapes : ouverture, question, réalisation de la requête, retour vers la question initiale et enfin, clôture.

– Une analyse des *actes de dialogue* (au sens de (Bunt, 1996) comme acte de langage contextualisé) met en évidence le fait que chaque énoncé du corpus peut être décomposé en segments auquel est associé un acte de dialogue. Une liste d'actes de dialogue a été établie en fonction des marqueurs discursifs relevés dans le corpus. Cette taxinomie, présentée dans (Loisel 2004), reprise et adaptée de (Weisser 2003), n'est pas présentée dans cet article.

– Une analyse des différentes Q-R. Le formalisme de QUD et GoDIS permet d'analyser précisément notre corpus en termes de Q-R selon le sous-dialogue courant.

– Une analyse de la cohérence : nous sortons QUD du cadre de la théorie sémantique et Godis du modèle de Dialogue H-M simple, pour les utiliser dans le cadre plus riche du dialogue H-H.

**4.1. Analyse sociolinguistique**

Cette analyse, en étudiant finement les interactions entre les deux interlocuteurs (expert et demandeur), montre tout le travail discursif mené durant l'entretien.

4.1.1. *Amorcer*

La première préoccupation des experts est d'amorcer le dialogue avec la machine.

*Expert : voilà, donc je vous écoute...*
*Expert : ok c'est parti / donc quelle question vous intéresse ?*

Les experts se font médiateurs et mettent en oeuvre des stratégies de réponses similaires, quoique menées sur des tempos différents : dans un cas, le cheminement est privilégié, tandis que dans l'autre, c'est une marche rapide vers la résolution du problème. Dès lors que la phase d'amorce, bien repérable, est lancée, les interactions suivent deux chemins divergents : soit l'expert estime que la question est adaptée et valide la demande, soit elle ne lui semble pas appropriée et elle donne alors lieu à un certain nombre de reformulations.

Dans la première situation, des marqueurs forts de validation « d'accord », des appréciatifs « bon », des connecteurs de conclusion « donc », indiquent que la question est valide et ne sera de fait pas remise en cause. Néanmoins, malgré la validité estimée de la question, il peut s'avérer nécessaire de préciser les choses :

*Expert : alors je vais écrire ce que vous êtes en train de me dire / j'aurais aimé avoir des renseignements sur les dons d'organes / donc j'écris ce que vous me dites et est-ce que éventuellement vous pouvez préciser un petit peu ou vous voulez qu'on parte comme ça / des types de renseignements c'est peut-être un petit peu vague*

Dans la deuxième situation, si la requête ne semble pas adaptée et vouée à un échec probable, l'expert se garde bien de refuser la question afin de maintenir le dialogue, mais invite en douceur l'enquêté à une reformulation de sa demande.

*Expert : alors on va essayer de préciser un peu cette question /--/*

### 4.1.2. *La gestion de l'interaction*

Une série d'itérations vient assurer la continuité de l'entretien. Lorsque le nombre de documents obtenus avec la requête est trop grand ou trop faible, un retour à la page d'accueil s'effectue :

*Expert : / bon ça ne va pas / on va revenir / on va essayer de voir*
*Expert : / dermatologie / et donc là je relance la recherche / donc / ce qui est quand même une maladie rare a priori ou alors des poux dans des cils*

Ce tâtonnement, récurrent tout au long des entretiens, donne lieu à des énoncés à la fois axiologique « c'est intéressant ça », et fortement modalisés « peut-être qu'on ne va pas trouver ».

L'expert prévient son interlocuteur par des modalisateurs épistémiques : « peut-être », « si jamais », des séquences comme « on verra bien », le verbe « essayer ».

*Expert : on peut peut-être essayer de voir si on peut affiner un tout petit peu les choses ou est-ce qu'on s'amuse à lancer la requête comme ça*
*Expert : on va peut-être essayer avec chauve et dermatologie /.../ on va lancer la recherche on verra bien ce que ça donne /.../ on peut rajouter des choses après donc faisons quelque chose de très général et puis on essayera de cerner des choses d'un peu plus près*

Ce tâtonnement nécessite d'informer l'interlocuteur que la requête pourrait conduire à un échec, mais dans le même temps, il est essentiel de le rassurer pour poursuivre l'interaction :

*Expert : je ne suis pas sûre que vous sortiez avec une réponse / dommage / non / non rassurez-vous / on va y arriver / on va y arriver / euh*

L'expert précise l'organisation de sa démarche, tentant d'inclure son interlocuteur dans sa recherche d'information. Le balancement entre le « nous » inclusif et le « je » est à ce propos significatif :

*Expert : je relance la recherche / donc nous n'avons toujours rien trouvé*
*Expert : donc nous lançons la recherche telle quelle / et puis / bien nous sommes toujours bredouille / donc nous revenons parce que nous avons toujours zéro ressource / nous revenons à la page précédente / je vais enlever examen*

Le « je » du locuteur prend par instant le dessus. Dès qu'il en prend conscience, le « on » ou le « nous » est vite rétablit.

La reprise des mots du demandeur avec la marque d'acquiescement signale le passage à l'acte d'écriture de la requête, mais montre dans le même temps la prise en compte de la parole de l'autre. « OK », « oui », « donc » sont autant de façon d'affirmer un but commun, celui de trouver une réponse à la question du demandeur. De la même façon, l'usage de l'impersonnel « on » et le verbe « préciser » marquent une reprise qui vise à mettre l'accent sur une tâche à mener en commun :

*Expert : On va essayer de préciser un peu cette question*

Ces marqueurs discursifs (marques d'approbation, d'interrogation ou de refus) guident l'interaction. Même si, de façon générale, le demandeur reste extrêmement laconique, l'ensemble de ces procédés montre à quel point les processus conversationnels sont fondés sur un principe de la coopération.Tout au long des entretiens, les experts essaient de garder le contrôle du dialogue en tentant de ne pas perdre la face, face à une machine qui ne répond pas toujours aussi bien qu'ils le souhaiteraient. Cela passe par des demandes de répétition, de confirmation ou de validation nécessaires en cas d'ambiguïté, voire de contestation. Tout un dispositif est donc mis en oeuvre pour assurer la réussite de l'interaction. Mais c'est essentiellement autour des termes à utiliser que les négociations discursives se mettent en place.

4.1.3. *La manipulation terminologique*

Dès le début des entretiens s'opère une négociation discursive qui vise à contraindre la terminologie utilisée par le demandeur, formatant celle-ci vers la terminologie CISMeF. Deux stratégies sont alors mises en place par l'expert, la première consistant en un balayage systématique de la terminologie CISMeF, la seconde conduisant à un formatage de la parole du demandeur.

– Balayage de la terminologie CISMeF

Le sens référentiel présuppose ici des stratégies de contextualisation communes. CISMeF offre des menus déroulants qui vont d'emblée servir aux experts de première base de travail. La solution est l'exhaustivité pour tenter de cerner au mieux le « bon mot » :

*Expert : bon on va regarder tout ce qu'on a*
*Expert : alors déjà je vais essayer de trouver un mot clé qui va pouvoir correspondre / alors on va chercher angoisse par exemple*

Le « par exemple » révèle la volonté de l'expert de rechercher un consensus : il faut que les deux protagonistes s'entendent sur les termes à utiliser. Mais parfois la méthode mène à l'échec :

*Expert : allons-y / voyons ce que ça donne / nous n'avons rien / c'est décidément assez désespérant*

Devant la difficulté, l'expert réclame des commentaires au demandeur :

*Expert : donc ça n'est pas / vous avez le droit de commentez / bien au contraire / je serais ravie que vous commentiez*

Si le demandeur intervient parfois de son propre chef, venant en aide à l'expert en lui suggérant d'ajouter une information, c'est le plus souvent l'expert qui impose son choix ; quand bien même l'expert sait laisser une place au dire de l'autre :

*Demandeur : ou les cheveux*
*Expert :à moins qu'on mette cheveu peut-être effectivement vous avez raison/cheveu allons-y*

Cette phase d'interprétation contextuelle, fournie en commentaires épilinguistiques, révèle comment l'outil s'explore.

– Le formatage de la parole du demandeur

Dès les premiers mots échangés, l'expert cherche à borner la question du demandeur, comme le démontre les nombreux marqueurs discursifs de reformulation, ou le recours à des catégories proposées par la base. Il arrive cependant que le mot utilisé par le demandeur soit d'emblée reconnu comme terme CISMeF :

*Expert : glaucome devrait être répertorié / il l'est tout à fait*

Le mot « glaucome » est identifié comme terme technique, autrement dit comme mot au statut spécifique dans la communauté médicale et, dès lors, apte à être utilisé. Le constat est immédiat : « il l'est tout à fait ».

Mais le plus souvent, un ajustement est nécessaire avant d'arriver à faire correspondre les mots utilisés par le demandeur avec ceux de la base, ce que signale l'embrayeur « alors déjà » dans l'extrait suivant :

*Expert : alors déjà je vais essayer de trouver un mot clé qui va pouvoir correspondre*

Cet ajustement est parfois laborieux. Dans cet autre extrait, les indices d'hésitation et les questions marquent la difficulté qui se présente à l'expert :

*Expert : alors là dans notre cas ce serait / euh: / ils mettent pas psychologie / donc on va faire sans / alors dans les qualificatifs donc comment venir à bout euh:: / donc on va essayer de trouver / euh:/ comment traduire ça / ce serait finalement des thérapeutiques*

L'expert a deux postures successives. Dans la première, il abandonne : « donc on va faire sans », tandis que dans la seconde, il se résout tant bien que mal (résignation marquée par l'adverbe « finalement ») à faire entrer les paroles de l'autre sous le terme « thérapeutiques ».

Loin d'être une « traduction » des mots du demandeur en termes CISMeF, nous assistons plutôt à une « réduction », à un formatage de la parole de l'enquêté qui voit ses mots contraints par la terminologie proposée.

– La manipulation terminologique.

En fonction des réponses données, l'interaction se reconfigure pour tenter d'aboutir à une réponse. Si les négociations terminologiques doivent *a priori* assurer le succès de la requête, tel n'est pas toujours le cas :

*Expert : donc on va réessayer / de faire une deuxième requête pour troubles anxieux /--/ /--/ hop /--/ ça par contre /--/ ça c'est pas terrible /*

L'évaluatif « c'est pas terrible » révèle un essai infructueux : le bricolage terminologique n'a pas été suivi de l'effet escompté. Une interprétation discursive est nécessaire, orientant tout un cheminement de reformulations.

En fonction des premières réponses de la base, les experts effectuent des sauts terminologiques mineurs, prédictibles par analogie sémantique. De même, il peut s'agir tout simplement de trouver l'hyponyme qui correspond plus précisément à la question de l'enquêté. Dans l'exemple suivant, il faut trouver « le bon » cholestérol dans l'ensemble des cholestérols :

*Expert : donc je note un résumé de la requête initiale donc cholestérol et traitement / / donc on va lancer d'abord / donc on a tous les mots clés CISMEF / cholestérol en fait partie naturellement / (rire) / ça c'est /.../ on a cholestérol HDL et cholestérol alimentaire /.../ il y a le bon / mais en fait c'est un taux très anormal de HDL / HDL d'accord / on va d'abord regarder ce qu'on nous propose dans cholestérol / on a 19 ressources / donc on va affiner ça //--/ on va voir le cholestérol HDL /--*

Le saut peut être aussi morphologique, lorsque l'expert passe de « cheveu » à « calvitie » par exemple.

Mais lorsque la question est « Comment venir à bout des crises d'angoisse ? », les choses se compliquent singulièrement. « Troubles anxieux » et surtout « thérapeutiques » induisent en effet un *bond terminologique* bien plus conséquent. La progression terminologique entre la phraséologie de départ et les termes d'arrivée est extrêmement sinueuse. Le problème est alors précisément dans ce passage de la périphrase « comment venir à bout ? » au terme tapé « thérapeutique », dont le syntagme donné comme synonyme « prise en charge » est bien éloigné au niveau morphosyntaxique et ne semble guère partager de traits avec le syntagme verbal « venir à bout de quelque chose ».

S'inscrivent donc dans l'espace de l'écran des zones que les protagonistes doivent remplir tant bien que mal. Tout le travail de négociation discursive mis en évidence lors de l'analyse montre que cet ajustement dépasse une simple adéquation entre équivalents sémantiques.

### 4.2. Analyse des sous-dialogues

Il est possible de décrire la structure globale des dialogues par des enchaînements de différents types de sous-dialogues, comme le montre la Figure 1. Les flèches pleines entre sous-dialogues représentent les transitions naturellement induites par la tâche. Les flèches pointillées représentent une relation de décomposition des sous-dialogues.

Un dialogue débute par une phase d'« ouverture », phase d'établissement du dialogue qui peut être aussi bien longue que courte. Elle consiste à accueillir l'interlocuteur, présenter CISMeF, et négocier la tâche à réaliser en commun. Il s'ensuit une phase de « choix de sous-dialogues », qui dans la plupart des cas, conduit à une phase principale de « recherche d'informations ». Celle-ci est composée d'une ou plusieurs séquences de recherche qui aboutiront chacune à la résolution d'un problème posé par le demandeur. Enfin, le dialogue se termine par une phase de « clôture » repérable par des marqueurs discursifs stables. Cette clôture est faite par l'un des interlocuteurs, lorsqu'il estime que le nombre de documents obtenus est raisonnable.

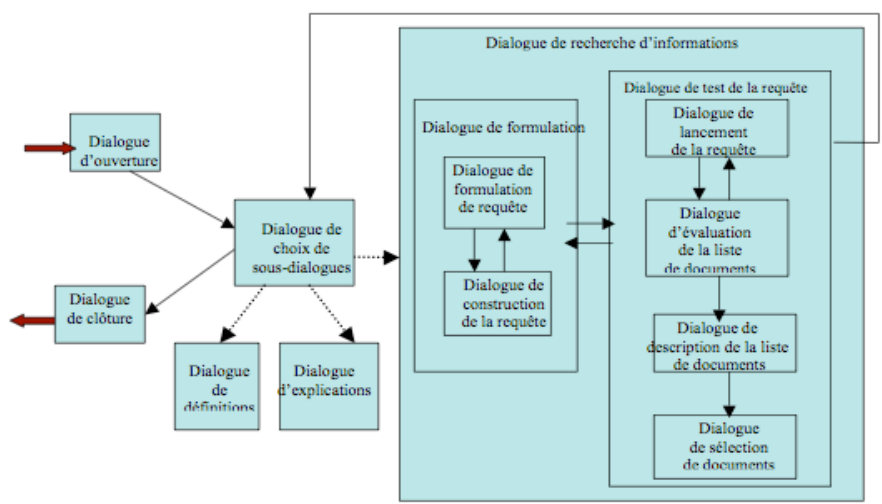

**Figure 1 : Enchaînements des différents sous-dialogues**

La séquence de recherche se décompose en deux parties : un sous-dialogue de formulation de la requête qui enrichit une requête courante et un sous-dialogue de test de la requête. A chaque instant de cette phase, les interlocuteurs peuvent revenir sur la requête courante.

La formulation de la requête se décompose en deux sous-dialogues : un premier sous-dialogue de formulation dont le but est de cerner le type de la demande. Un certain nombre de reformulations et de précisions s'effectuent de part et d'autre, qui ont pour objectifs tout à la fois de borner le thème et de préciser les termes de la question. Dans un second sous-dialogue de construction de la requête, un travail de reformulation se déploie. Durant cette phase, la requête est élaborée en coopération avec l'usager. Les termes constituant la requête sont discutés un à un en adéquation avec la terminologie CISMeF.

Suit une phase de test de la requête obtenue précédemment. Plusieurs requêtes sont exécutées lors du sous-dialogue de lancement de la requête et les résultats sont présentés au demandeur dans un sous-dialogue d'évaluation (quantitative) de la liste des documents, en fonction du nombre de documents trouvés. Cette liste est alors décrite qualitativement au demandeur, dans un sous-dialogue de description de la liste de sous-documents et un ou plusieurs documents peuvent être analysés en détail dans un dialogue de sélection de documents. A tout moment, cette phase de test de la requête peut être interrompue par des demandes de précisions.

En plus des demandes de recherche de documents, les corpus font apparaître des demandes de définition de termes médicaux et d'explications sur le système lui-même. Tous ces sous-dialogues ont été observés dans nos corpus, mais nos corpus ne les font évidemment pas apparaître tous à la fois.

*4.2.1. Sous-dialogues : segments à but épistémique ou actionnel*

Nous considérons à la suite de (Vernant, 1997) et de (Xuereb, 2004) que les buts des sous-dialogues peuvent être situés sur deux axes : épistémiques et actionnels.

– Buts épistémiques : la résolution du segment dialogique permet d'apporter des connaissances sur un terrain commun. Nous appelons les sous-dialogues « sous-dialogues de Q-R » puisque l'entrée dans celui-ci se fait lorsque la question est posée et il se conclut lorsqu'une réponse est apportée ou que la question est abandonnée.

– Buts actionnels : la résolution du segment dialogique permet d'accomplir une action conjointe (saluer, lancer une requête, etc.). Le segment dialogique se conclut lorsque l'action est réalisée ou abandonnée. Nous appelons ces sous-dialogues « dialogues d'actions ».

Les relations entre deux segments dialogiques *A* et *B* (qui peuvent être indistinctement épistémiques ou actionnelles) sont caractérisées ainsi :

– *Dominance(A, B)* : résoudre *B* est nécessaire pour résoudre *A*,

– *Pré-séquencement(A, B)* : résoudre *A* est nécessaire avant de résoudre *B*,

– *Séquencement(A, B)* : résoudre *A* est généralement effectuée avant *B*. Cette relation signifie qu'un interlocuteur préfère sans autre indication s'engager dans le but *B*, juste après le but *A*.

Exemple :
```
PréSéquence(Ouverture, SéquenceRecherche)
PréSéquence(SéquenceRecherche, Clôture)
```

Les sous-dialogues sont eux-mêmes constitués de segments dialogiques que nous appelons Q-R et segments d'action.

*4.2.2. Phénomènes d'interruption dialogique*

La structure linguistique des dialogues de nos corpus présente un certain nombre de phénomènes que nous qualifions d'opportunistes. En effet, cette structure est constamment interrompue par un ensemble de phénomènes dialogiques à l'initiative de l'un ou l'autre des interlocuteurs qui suspendent les sous-dialogues. Ces phénomènes ne peuvent pas être planifiés et illustrent bien les difficultés d'une analyse uniquement centrée sur la planification.

Certains des sous-dialogues peuvent être jugés facultatifs. Certains même se déroulent de manière silencieuse, de manière implicite. Dans certains cas, l'action sous-jacente est réalisée, mais ne donne pour cette action aucun dialogue. C'est par exemple le temps de réflexion et l'ensemble des actions de recherche réalisées qui permettent de dire que l'action sous-jacente est effectuée. Ici, il n'y a pas de rupture : seule la forme linguistique du dialogue semble incomplète. Le sous-dialogue présentant ce phénomène le plus fréquemment est le sous-dialogue de lancement de la requête. Il peut y avoir aussi absence de salutations au début d'un dialogue ; l'intention d'ouverture du dialogue est bien réelle, mais elle est sous-entendue.

Chacun de ces sous-dialogues peut être annulé explicitement. Plus particulièrement, dans le corpus, le sous-dialogue de test de la requête peut être interrompu pour chacun de ses sous-dialogues, mais l'annulation peut porter sur une séquence de recherche complète. En cas d'annulation, le dialogue se retrouve soit dans le sous-dialogue précédent, soit dans le sous-dialogue de choix.

Chacun des sous-dialogues peut être annulé implicitement par l'un des interlocuteurs qui revient alors dans le sous-dialogue précédent. C'est en particulier le cas lorsque le sous-dialogue de test de la requête est annulé. Il faut différencier ce cas du précédent où le sous-dialogue est accompli, mais de manière silencieuse. Par exemple, le demandeur propose un nouveau terme pendant que l'expert effectue le test de la requête. Dans ce cas, le dialogue revient dans un état de sous-dialogue de formulation de la requête.

Des sous-dialogues incidents (Luzzatti 1989) peuvent apparaître n'importe où afin d'assurer le partage du fond commun.

Des sous-dialogues de choix peuvent apparaître lorsqu'il y a ambiguïté dans le choix d'un sous-dialogue ou pour relancer la recherche. Il y a trois points de choix dans cette tâche de recherche d'informations : passer du test de la requête vers un sous-dialogue de formulation pour compléter la requête, passer d'une séquence de formulation de requête vers une nouvelle recherche ou clore le dialogue.

Enfin, des sous-dialogues incidents peuvent être emboîtés pour réaliser des digressions. Dans ce cas, le dialogue régissant reprend à l'endroit où il a été laissé. Il s'agit de stratégies dialogiques collaboratives (Caelen 2003).

**4.3. Analyse des différentes Q-R**

A chaque sous-dialogue correspond un ensemble de Q-R qui font progresser la tâche de construction de la requête dans la terminologie CISMeF. Dans cette section, nous prenons l'exemple des sous-dialogues de formulation libre de requête ainsi que ceux de construction de la requête.

*4.3.1. Sous-dialogues de formulation libre de requête*

Nous n'analysons pas ici toutes les questions sous-jacentes à la formulation de la requête. La sémantique de toutes ces questions est très complexe et nous supposons que pour l'expert et pour le système à concevoir, ces élaborations ne sont qu'un moyen d'obtenir des termes CISMeF sur lesquels travailler. Pendant cette phase, les actes d'établissement sont très fréquents et permettent de représenter les différents énoncés.

La formulation initiale du demandeur peut se faire par un acte de dialogue de requête sous la forme d'une question :

*Demandeur : ma question est comment venir à bout des crises d'angoisse ?*

```
Answer(RequêteInitiale(« comment venir à bout … ?»))
```

Elle peut aussi se faire par un acte de dialogue indirect comme : « je voudrais savoir », « j'aimerais savoir », « je suis intéressé par » :

*Demandeur: donc bonjour, moi je suis intéressé par tout ce qui est problème de cholestérol*
```
Greet, Answer(RequêteInitiale(« pb cholestérol »))
```

La formulation initiale peut encore se faire sous forme de réponses courtes :

*Demandeur : par rapport aux articulations en fait / douleurs articulaires et autres*
```
ShortAnswer(RequêteInitiale(« articulations »)
ShortAnswer(RequêteInitiale(« douleurs articulaires »)
```

Les demandeurs peuvent ajouter spontanément des élaborations de leurs requêtes. Pour exprimer ces relations sémantiques, il est nécessaire d'utiliser un formalisme de représentation sémantique plus complet (comme la SDRT, Segmented Discourse Representation Theory (Asher 1993)), mais dans notre étude, nous isolons uniquement les termes de la requête qui nous intéressent dans ces actes de dialogues subordonnés.

*Demandeur : donc j'ai une amie qui est épileptique et je voudrais savoir quel est le risque pour la grossesse.*
```
Answer(RequêteInitiale(« épileptique »)
Answer(RequêteInitiale(« risque pour la grossesse »)
```

L'expert peut demander des précisions. Une fois le sous-dialogue de formulation de la requête initiale terminé, cette demande de précision permet d'élaborer la requête dans la terminologie CISMeF. Le passage d'une étape à l'autre est très souvent signalé par l'expert par un acte de dialogue InformIntent.

*Expert : on va essayer de préciser tout ça /*
```
InformIntent(Action(precision))
```

Cet acte de dialogue d'information, s'il est accepté, permet soit de partir sur un ajout de requête, soit d'entrer dans un nouveau sous-dialogue de construction de la requête.

Il peut aussi exister des demandes de précision générale permettant à l'expert d'obtenir des précisions sur des termes de la requête initiale. Il s'agit d'obtenir de nouveaux termes pour préciser la recherche à partir des termes déjà obtenus.

*Expert : Donc ensuite on va essayer de trouver euh / qu'est-ce que vous entendez par problèmes avec la nourriture ?*
*Demandeur : euh / la relation qu'une personne peut entretenir vis-à-vis du fait d'ingérer des aliments / donc typiquement j'ai pensé à l'anorexie mais bon /*
```
E : RequestInfo(?λT.PrécisionsGénérales(T,« pb avec la nourriture »))
```

*4.3.2. Sous-dialogues de construction de la requête*

La construction de la requête se fait soit de manière informelle, sans citer explicitement les termes de la terminologie CISMeF (mots-clés, qualificatifs, etc.), soit de manière explicite, lorsque la terminologie a été présentée au demandeur.

Ajout ou proposition de mots-clés

L'expert peut ajouter des mots-clés, sans demander de confirmation à son interlocuteur, grâce aux connaissances médicales du moteur de recherche lui-même.

*Expert : Déjà dans les mots-clés / on va rentrer épilepsie*
```
Inform(AjoutMotClé(epilepsie.mc))
```

L'expert ou le demandeur propose un mot-clé à partir des termes employés (qui ne sont pas forcément des mots-clés), en utilisant des analogies ou en recherchant dans la terminologie un mot-clé adéquat.

*Expert : On parle apparemment de troubles anxieux*
```
Suggest(AjoutMotClé(troubles anxieux.mc))
```

*Demandeur : Cherchez sur tendinite peut-être /…/*
*Expert : Chercher sur euh, tendinite oui /*
```
D : Suggest(AjoutMotClé(tendinite.mc))
E : icm:sem*pos(AjoutMotClé(tendinite.mc))
    icm:acc*pos(AjoutMotClé(tendinite.mc))
```

Lorsqu'une requête n'est pas satisfaisante, il est nécessaire de trouver d'autres mots-clés pour préciser ou élargir celle-ci. Ceci se fait grâce à plusieurs types de Q-R telles que la recherche d'hyperonymes ou de synonymes ou la combinaison de termes.

*Expert : ensuite on regarde tous les mots-clés qui sont là-dedans / il nous donne un terme plus générique / c'est les arthralgies / donc maladies des articulations*
```
Suggest(AjoutMotCléHyperonyme(gonarthrose.mc, arthralgie.mc))
```

*Demandeur : euh / par contre c'est bizarre, on n'a rien d'autre / il n'y a pas un synonyme de cette maladie ?*
```
D : RequestInfo(?λTermeSynonyme(maladie1, Terme))
```

*Expert: On va essayer ce mot avec l'autre*
```
Inform(Combinaison(Mot1, Mot2))
```

Suppression de termes de la requête

Une Q-R permet également dans certaines circonstances de supprimer un terme de la requête courante (pour élargir la requête et obtenir plus de résultats).

*Expert : oui / alors on va enlever ce mot-clé là / (« ce mot clé là » = anorexie.mc)*
```
inform(SuppressionTermeRequete(anorexie.mc))
```

*Demandeur : mais peut-être que c'est trop spécifique ? (« c' » = anorexie.mc)*
```
Suggest(SuppressionTermeRequete(anorexie.mc))
```

### 4.4. Analyse de la cohérence des sous-dialogues

Après avoir passé en revue les différentes Q-R du corpus, nous définissons leurs relations de cohérence.

*4.4.1. Instanciations des Q-R*

Nous nous sommes intéressés aux marqueurs de cohérence les plus fréquents dans le corpus comme les conjonctions et adverbes, permettant d'expliciter certaines relations monologiques ou dialogiques entre Q-R du corpus. Certains adverbes et conjonctions n'ont pas le rôle de connecteur, cependant la plupart de ces mots indiquent des relations entre Q-R. Ils ne constituent toutefois qu'un indice sur la relation et ne doivent être considérés que comme tels. Les relations identifiables sont les suivantes :

– la progression sur l'axe régissant (relation intentionnelle) « donc », « alors » ;

– des marqueurs de séquence, satisfaction/précédence « après », « ensuite », « puis » et leur variante « alors après » et « alors ensuite » ;

– nouveau sous-dialogue, première réponse à une question « déjà » ;

– des marqueurs de sous-dialogue collaboratif « par exemple », « parce que », « puisque » ;

– des marqueurs de stratégies coopératives « si », « peut-être » ;

– « aussi » et « encore » permettent de préciser une nouvelle réponse à une question, « par contre » permet de repréciser une réponse en adjoignant une relation de contraste. Ces relations sont à la fois intentionnelle et sémantique ;

– « plutôt » indique la correction d'une réponse à une question ;

– « sinon » permet au contraire d'abandonner une question ;

– d'autres relations sémantiques qui ont à la fois valeur intentionnelle et sémantique « mais », « quand même », « toujours ».

*4.4.2. Plusieurs réponses satisfaisantes*

Pour certaines questions, plusieurs réponses sont possibles. La première réponse est donc résolvante, mais ne clôt pas la question.

*Expert : Est-ce qu'éventuellement vous pouvez préciser un petit peu ?*
*Demandeur : Bah savoir les démarches à accomplir si on veut être donneur d'organes par exemple /…/ S'il y a des examens à passer*

Il y a sur cet exemple deux réponses satisfaisantes à la question posée par l'expert. L'utilisation de connecteurs comme « aussi », « encore » permet de préciser qu'il faut effectivement une nouvelle réponse à une question.

GoDIS simplifie ce problème en considérant qu'une seule réponse est acceptable si elle est potentiellement résolvante. Clairement ici, les deux réponses sont acceptables et résolvantes. D'ailleurs, la théorie des questions en discussion précise bien qu'il doit y avoir relativisation pragmatique : c'est toujours en rapport aux buts des interlocuteurs qu'une réponse peut réellement être résolvante. Cependant, d'autres questions n'admettent qu'une seule réponse.

*Demandeur : Rhumatologue / rhumatologie dans la première /*

*Expert : D'accord / euh, donc on va essayer ça / avec rhumatologie / donc je lance la recherche comme ça /*

Dans l'exemple ci-dessus, le demandeur propose une réponse à la question qui est acceptée, mais qui ne correspond pas à un métaterme CISMeF. Le demandeur répond alors par un nouveau métaterme « rhumatologie » sans que la question ne lui soit reposée. Ici, la première réponse est résolvante, mais l'utilisateur peut toujours proposer une seconde réponse, un autre métaterme. Donner une deuxième réponse à la question doit alors être interprété comme une correction de la première réponse.

*4.4.3. Information, suggestion et stratégie coopérative*

– Réponses à la question données par le questionneur lui-même

Dans certains cas, il est difficile de savoir si une question est de cette forme car l'expert met un certain temps avant de répondre à sa propre question. Le demandeur pourrait interpréter cela comme une réelle requête d'information et y répondre. Dans QUD et GoDIS, le fait de poser une question ne précise pas qui doit répondre à cette question. Le fait que ce soit le questionneur qui réponde à sa propre question ne modifie pas le mécanisme.

D'un autre point de vue, le fait que l'expert réponde à sa propre question peut être interprété comme l'adoption d'une stratégie coopérative afin d'aider le demandeur. Ici encore, le demandeur a la liberté de contester cette réponse et d'en proposer une autre (phénomène de ré-accommodation).

*Expert : Comment traduire ça ? Ce serait finalement des thérapeutiques*

– Questions présentées sous forme de suggestions

Les suggestions permettent à l'expert de proposer lui-même une réponse à une question sans même la poser au demandeur. La présentation sous forme de suggestion permet de demander l'approbation du demandeur.

– Questions présentés sous forme d'`Inform`

Une troisième forme de stratégie coopérative consiste à proposer des réponses à l'interlocuteur sans même lui demander son approbation.

*4.4.4. Accommodation de questions et d'actions*

Nous pouvons identifier deux mécanismes d'accommodation qui permettent d'interpréter certains actes de dialogues indirects.

– Accommodation de questions polaires vers des questions ouvertes

Cette accommodation se fait lorsqu'une question à réponse polaire est mise en discussion. Dans ce cas, une question ouverte implicite est également mise en discussion. Cela permet d'expliquer le fait que l'interlocuteur peut répondre à cette question par une réponse typée.

*Expert : Est-ce qu'il y a une autre question pour étendre la recherche ?*
*Demandeur : chercher sur tendinite peut-être / --/ /--/*

– Actes de dialogue indirects entre questions et actions

Il existe également un mécanisme d'accommodation entre une question et une action associée.

*Demandeur : Donc je crois qu'on va en rester là non ?*
`RequestInfo(?Clôture)`
*Expert : D'accord ça marche /*
`icm:acc*pos(Action(Clôture))`

Comment interpréter qu'une réponse à une question n'est pas un acte de dialogue `Answer` ? Poser la question `?clôture` propose une action équivalente de clôture qui est ensuite résolue par l'acte de dialogue d'acceptation. Ainsi, si une question polaire est ajoutée et qu'il existe une action associée, alors cette action est également ajoutée.

*Expert : Est-ce que tu veux préciser autre chose que ça ou euh ?*
*Demandeur : Non*
*Expert : Non ? donc a priori on n'a pas trouvé de chose précisément là-dessus / mais plus sur les troubles de sommeil en général /*

Enfin cet exemple illustre le phénomène de refus et le phénomène de question implicite. Lorsque la question principale `RequestInfo(?Précisions)` est annulée, la question implicite `RequestInfo(?λx.Précisions(x))` est également annulée.

### 4.4.5. Mise en terrain commun des Q-R

Les exemples tirés du corpus confirment l'hypothèse de (Allwood 1992) qu'un acte de dialogue de niveau compréhension sémantique est généralement suivi d'un acte d'acceptation. C'est notamment le cas dans la formulation initiale de la requête, pour être sûr que tous les mots recherchés sont bien intégrés.

*Demandeur : Bah savoir les démarches à accomplir si on veut être donneur d'organes par exemple*
`Answer(RequêteInitiale(« Demarches », « donneur d'organes »))`
*Expert : Donc savoir les démarches en tant que donneur d'organes, d'accord*
`icm:und*pos(RequêteInitiale(« Demarches »,« donneur d'organes »))`

A notre connaissance, les recherches précédentes ne traitent que des acceptations sur des questions. Or, nous trouvons dans notre corpus des demandes explicites d'acceptation sur l'ensemble d'un sous-dialogue.

*Expert : Ca va / ça vous plaît ?*
`RequestInfo(?icm:acc*chk(Action(RequêteInitiale)))`
*Demandeur : Oui / ça ne me paraît pas hors de propos / Tout va bien /*
`icm:acc*pos(Action(RequêteInitiale))`

### 4.4.6. Relations entre Q-R

Dans notre corpus, les relations entre Q-R sont interactionnelles, sémantiques et intentionnelles d'une part, subordonnées et coordonnées d'autre part, en reprenant la classification établie par (Prévot, 2004) et approfondie par (Xuereb, 2005). Nous nous sommes concentrés sur les relations intentionnelles et les mécanismes permettant de passer d'une Q-R à une autre en progressant dans la tâche.

Les relations intentionnelles permettent de décrire des liens entre questions, entre la tâche et le dialogue.

- Relation de dépendance entre Q-R ou entre Q-R et action

Dans notre corpus, il y a de nombreuses relations de dominance entre les actions de haut niveau liées à un sous-dialogue et des questions. Par exemple, l'action de haut niveau de construction de requête `Action(ConstructionRequête)` domine la question d'ajout de mots-clés `?λM.AjoutMotClé(M)`.

- Relation de satisfaction-précédence

Cette relation est indispensable pour modéliser notre corpus. Les différentes actions et questions déterminent les différents sous-dialogues. Ainsi, il y a une relation de satisfaction-précédence entre `Action(ConstructionRequête)` et `Action(LancementRequête)` dans la mesure où la construction d'une requête valide est indispensable à son action de lancement.

- Pseudo-relation de séquence et accommodation globale

Lorsque deux questions sont dominées par une Q-R principale, mais qu'il n'y a pas de relation de satisfaction-précédence, elles sont séquencées dans un ordre plus ou moins arbitraire. Cet ordre est dirigé par l'interlocuteur guidant l'échange. Ainsi, dans notre corpus, un des experts utilise peu de stratégies coopératives : il pose une question pour ajouter des mots-clés, puis une autre pour ajouter des qualificatifs, puis une autre pour les métatermes. Cette séquence est révisable et si le demandeur choisit d'ajouter les termes dans une séquence autre que celle proposée, les réponses doivent être accommodées par accommodation globale.

*4.4.7. Transitions entre sous-dialogues*

Les jeux de dialogue, comme dans (Maudet, 2001), proposent de décrire les transitions entre deux jeux de manière explicite. Dans notre corpus, un acte de dialogue correspond à chacune des transitions entre sous-dialogues.

– Demande d'ouverture explicite d'un sous-dialogue.

Cela permet de choisir de lancer un nouveau dialogue. Ce peut aussi être un choix simple comme l'exemple ci-dessous : la réponse négative signifie rester dans le sous-dialogue courant.

*Expert : Est-ce que vous avez une autre requête ?*
*Demandeur : Non, c'était tout.*

– Ouverture explicite d'un sous-dialogue

L'acte de dialogue `InformIntent` intervient comme un acte de dialogue d'ouverture qui introduit lui aussi une question par accommodation de fait sur le désir de l'interlocuteur d'entrer dans ce sous-dialogue. Si la réponse est négative, l'entrée dans le sous-dialogue sera refusée.

*Demandeur : Ah bah dites-moi / Attendez bougez pas je vais rapprocher*
*Expert : Non / mais j'essaie / Action chimique / Analyse*

– Demande de fermeture explicite du sous-dialogue

Le problème de l'établissement peut s'étendre sur un sous-dialogue entier ce qui signifie que la fermeture d'un sous-dialogue entier peut être demandée par l'acte de dialogue `?icm:acc*chk(Action(_))`.

L'acte permettant de fermer un dialogue est toujours absent de notre corpus. On peut le représenter par `done(Action(_))`. Il est souvent implicité par l'ouverture d'un nouveau sous-dialogue.

– Passage implicite grâce à une relation de séquence entre sous-dialogues.

La tâche est représentée par des actions liées par des relations de séquences et/ou de satisfaction-précédence. Il existe également des successions prototypiques de sous-dialogues comme, par exemple, le passage de la formulation de la requête initiale à l'élaboration de la requête.

*Expert : D'accord / je mets d'abord l'énoncé initial / donc c'est problème avec la nourriture ?*
*Demandeur : oui /--/*
*Expert : D'accord /--/ donc on va déjà aller dans les accès thématiques /*

Le passage à un nouveau sous-dialogue est marqué par l'utilisation de connecteurs comme « donc » ou « déjà » qui montrent que le dialogue est la première Q-R d'un nouveau sous-dialogue.

– Relance d'un sous-dialogue

Cela consiste simplement à repartir dans un dialogue lorsque celui-ci est terminé, éventuellement en le proposant à l'interlocuteur.

– Passage dans un sous-dialogue collaboratif (sous-dialogues emboîtés)

Les dialogues de stratégies collaboratives interviennent lorsqu'il y a des subordonnées coordonnées sémantiques, sans liens directs avec la tâche. Le plus souvent, il y a des relations d'aide, de clarification, d'explication du problème. Lorsque la séquence collaborative se termine, le contexte initial doit être retrouvé.

– Stratégies coopératives (sous-dialogues séquencés)

S'il y a un ordre prototypique de sous-dialogues, l'expert lui-même peut ne pas suivre cette séquence. Une transition par stratégie coopérative permet d'entrer ou de proposer d'entrer dans un sous-dialogue, sans suivre la relation de séquence, mais en respectant les autres relations intentionnelles.

*Expert : Est-ce qu'éventuellement vous pouvez préciser un petit peu ?ou vous voulez qu'on parte comme ça / des types de renseignements parce que c'est peut-être un petit peu vague*

Dans le dialogue précédent, l'expert propose au demandeur de choisir entre deux sous-dialogues, celui indiqué par la relation de séquence et un autre.

– L'accommodation de sous-dialogue

Elle permet de court-circuiter une phase de recherche.

*Expert : voilà / donc je commence la recherche avec ça (parasomnie) / ok / bon / alors il y a rien là-dedans*

*Demandeur : Donc je crois qu'on va en rester là non ?*

Le demandeur propose la fin de l'entretien après un échec. C'est le passage d'un sous-dialogue de description de documents à un sous-dialogue de fermeture. Cette suite est non attendue par l'expert, mais elle peut être accommodée car les relations de satisfaction-précédence entre sous-dialogues sont respectées. Le connecteur « donc » indique le changement de sous-dialogue.

**5. Modèle de l'agent dialogique Cogni-CISMeF**

Pour intégrer les éléments de l'analyse du corpus dans un agent dialogique Cogni-CISMeF, nous proposons une architecture modulaire comprenant trois composantes : le modèle de la langue, le modèle du dialogue et le modèle de la tâche. (Loisel 2008) présente une description complète de ces modèles.

– Le modèle de la langue reçoit la demande de l'utilisateur sous forme d'un énoncé en langue naturelle. Il permet de reconnaître un acte de dialogue appliqué à un contenu propositionnel pré-codé. Il réalise trois analyses de cet énoncé :

- une analyse sémantique qui décompose l'énoncé en données lexicales utilisées par les deux autres analyses pour la reconnaissance de thématiques et de termes CISMeF,

- une analyse pragmatique qui repose sur un interpréteur d'actes de dialogue,

- une analyse contextuelle qui convertit les représentations en une représentation propositionnelle compréhensible pour le gestionnaire de dialogue.

– Le modèle de la tâche encapsule directement l'interface CISMeF pour accéder à la base de documents médicaux. Il comprend aussi un constructeur de requêtes à partir des termes reconnus de la demande et un interpréteur de résultats permettant d'affiner la requête si nécessaire.

– Le modèle du dialogue comprend un gestionnaire du dialogue et un générateur de dialogues.

Le gestionnaire du dialogue modélise les sous-dialogues observés dans le corpus sous la forme d'une bibliothèque de plans et gère un *Etat d'Information* ainsi qu'une représentation du terrain commun. Il repose sur GoDIS auquel nous avons ajouté la modélisation des questions à réponses multiples, la modélisation de l'accommodation de questions et d'actions, l'ajout de relations d'accommodation pour la prise en compte des relations intentionnelles et enfin, l'ajout de stratégies dialogiques donnant de la souplesse au dialogue.

Le générateur de dialogues est fondé sur des phrases à trous qui permettent de produire des énoncés présentés à l'utilisateur.

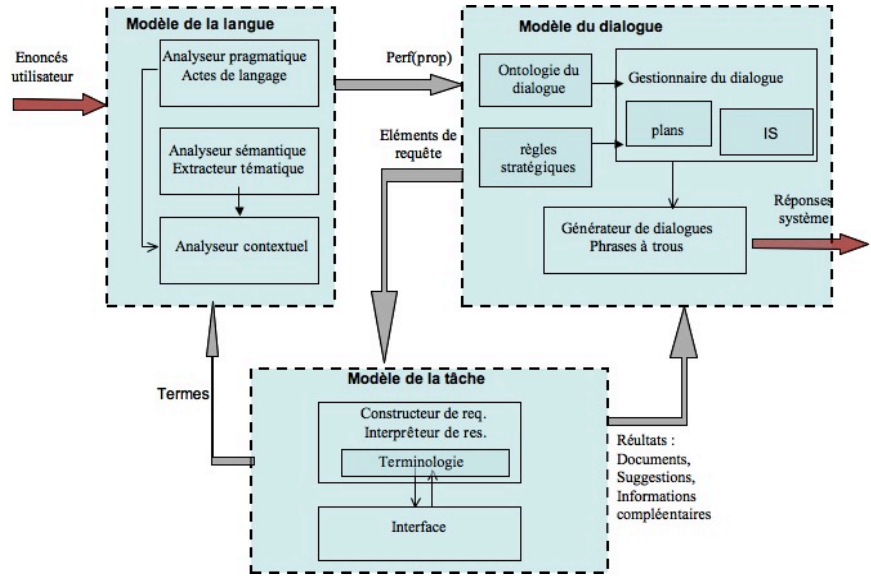

**Figure 2 : Modèle tri partite de Cogni-CISMeF**

En ce qui concerne l'implémentation, nous avons privilégié une approche pragmatique en réutilisant les outils de GoDIS. Chacun des modules (de la langue et de la tâche) comporte deux parties : une indépendante de l'IS (implémentée en Java) et une autre dépendante de l'IS (développée en Prolog).

Pour mieux comprendre le fonctionnement du système, la figure 3 montre le suivi d'un dialogue entre le système et un utilisateur en indiquant les règles de dialogue utilisées.

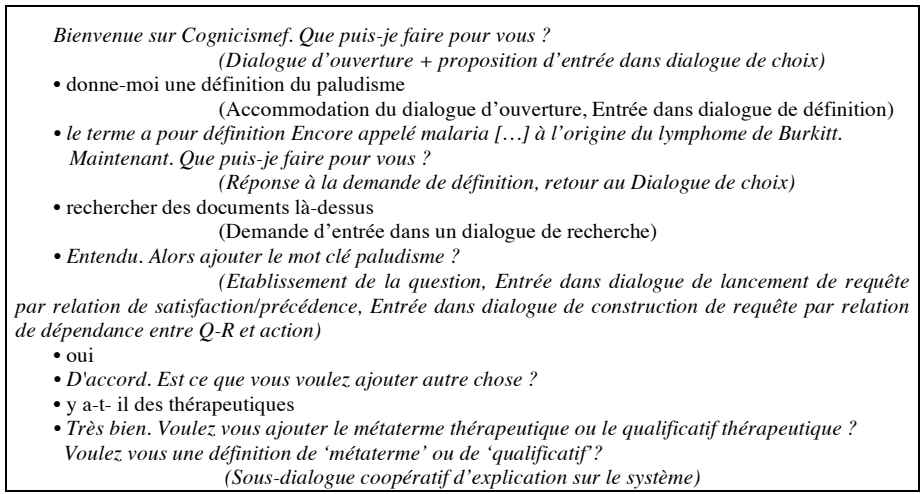

```
• Qu'est-ce qu'un qualificatif ?                            (Dialogue incident)
• Les qualificatifs sont des concepts généraux, qui peuvent être affiliés à un mot clé pour en
préciser le sens.
          • ajouter le qualificatif thérapeutique    (Retour au dialogue de construction de requête)
• Okay. Est ce que vous voulez ajouter autre chose ?
• non
• Entendu. c'est noté. Alors La requête a été ajoutée. Alors Voici la requête actuelle :
motcle(paludisme), qualificatif(thérapeutique). Pour l'instant, la requête vous convient-elle ?
              (Sortie du dialogue de construction de requête, Retour au dialogue de
lancement)
• oui
• Très bien. J'ai lancé la requête. J'ai mémorisé la requête j'ai trouvé des résultats. Alors Il y a
dans cette liste 11 documents. Il y a trop de documents. On peut essayer des termes plus spécifiques.
    [… Proposition de mots-clés plus spécifiques par expansion de requête, de qualificatifs associés,
…]
              (Succession de Stratégies collaboratives pour affiner la requête)
• Alors Etes vous patient ou médecin ?                      (Accommodation de question)
• patient
• On peut essayer d'ajouter des documents spécifiques pour les patients.
• oui             (Action associée à la question accommodée : type_ressource(patient))
• D'accord. Alors Voici la requête actuelle : Mots clés(paludisme, thérapeutique),
qualificatif(traitement médicamenteux), type_ressource(patient)
   Alors Pour l'instant, la requête vous convient-elle ?     (Mise en terrain commun)
• oui
• Okay. J'ai lancé la requête. J'ai mémorisé la requête j'ai trouvé des résultats. Alors Il y a dans
cette liste 1 documents. Alors Voici les titres de ces documents :
   1.   Modèle de chapitre pour les manuels PCIME - la prise en charge intégrée des maladies de
l'enfant
 • Alors Je viens de lancer l'évaluation. Alors Quel numéro de document souhaitez-vous ?
                   (Etablissement, Entrée dans dialogue de sélection de document)
• le premier
• le document sélectionné est :
Modèle de chapitre pour les manuels PCIME - la prise en charge intégrée des maladies de l'enfant
Une description du doc est prise en charge […] des enfants malades dans un petit hôpital ; 42p
Alors Est ce que l'on recherche autre chose ?
• non
• Entendu. Cette séquence de recherche est terminée. Alors Quelle question vous intéresse ?
```

**Figure 3 : Exemple de dialogue Cogni-CISMeF**

Pour finir notre étude, nous devons valider l'apport effectif de notre système de dialogue H-M. Son évaluation se fera selon le calcul de la plus-value apportée au système CISMeF. Pour les mêmes questions médicales émanant de notre corpus, il s'agira de comparer les requêtes construites par les demandeurs utilisant CISMeF, celles proposées par le conservateur et celles construites par Cogni-CISMeF suite à un dialogue avec les demandeurs. Nous mesurerons la plus-value (en termes de précision et de rappel) respective de Cogni-CISMeF et du conservateur par rapport à la requête seule de l'utilisateur dans CISMeF. Puis, nous étudierons les possibles différences en termes de précision et de rappel entre Cogni-CISMeF et le conservateur. Actuellement, le système n'est pas encore assez abouti pour une telle expérience. Il nécessite, au minimum, d'améliorer l'analyse sémantique en incorporant un lexique conséquent.

## 6. Conclusion et perspectives

Cet article présente une étude complète de la conception d'un système de dialogue H-M, du recueil de corpus à une implémentation. Nous avons appliqué QUD, une théorie du dialogue très peu traitée dans la communauté francophone et intégré l'application GoDIS. Le corpus de dialogues humains a été analysé sur différents niveaux, ce qui a permis de décomposer des dialogues en sous-dialogues liés à la tâche. L'analyse des dialogues en termes de Q-R décrivent des relations de cohérence entre sous-dialogues. Les Q-R suivent la tâche tout en offrant la souplesse des dialogues humains. En partant de QUD et de GoDIS, nous avons montré que ces théories, bien que moins avancées que des théories complexes comme la SDRT, sont des modèles implantables. Nous les avons enrichies par une sémantique des Q-R et l'utilisation de stratégies dialogiques.

Une grande couverture linguistique est indispensable pour tester en situation réelle une application de dialogue H-M. Pour notre approche, un problème de passage à l'échelle se pose pour augmenter cette couverture lexicale et permettre l'évaluation du système. Nous envisageons d'utiliser le système de reconnaissance de termes médicaux de (Pereira, 2008) qui intègre un plus grand nombre d'outils.

Notre gestionnaire de dialogue est capable de résolution de problèmes dialogiques non triviaux, mais tous les éléments observés dans l'analyse n'ont pas encore été intégralement implantés. Ainsi, l'analyse a montré que des actes de dialogues d'explications liés par des relations subordonnées sémantiques peuvent être proposés par l'expert. Dans notre implémentation, ces explications sont proposées uniquement en génération et ne sont pas prises en compte au niveau dialogique. Par exemple, l'utilisateur ne peut pas poser une question de type « pourquoi ? ».

Notre volonté de départ de prendre en compte le point de vue de l'utilisateur par la capture et l'analyse d'un dialogue en cours d'action, nous conduit à mesurer la complexité des phénomènes interactionnels inhérents à tout dialogue H-H. Le travail qui reste à accomplir est immense pour permettre, à l'instar d'une navigation sur une plate-forme de jeu, un réel couplage opérationnel entre l'utilisateur et une plate-forme informatique. Mais les conclusions de l'analyse linguistique du corpus, tout comme celles d'un système informatique de dialogue basé en premier lieu sur la question, ne sauraient rester vaines. Elles sont au contraire, pour notre équipe, les marqueurs indispensables de recherches pluridisciplinaires à venir.

## 7. Bibliographie